\documentclass[apjl]{emulateapj}
\usepackage{comment}
\usepackage{ifthen}
\usepackage{amsmath}
\usepackage{amsfonts}
\usepackage{amssymb}
\usepackage{multirow}
\usepackage{subfigure}
\usepackage{epsfig}

\usepackage[normalem]{ulem}
\usepackage{color}

\newcommand{\logg}{$\log g$}
\newcommand{\teff}{$T_{\rm eff}$}
\newcommand{\msun}{M$_\odot$}

\begin{document}

\title{Larger Planet Radii Inferred from Stellar ``Flicker" Brightness Variations
of Bright Planet Host Stars}
\author{Fabienne~A.~Bastien\altaffilmark{1},
Keivan~G.~Stassun\altaffilmark{1,2},
Joshua~Pepper\altaffilmark{3,1}}

\altaffiltext{1}{Vanderbilt University, Physics \& Astronomy Department,
1807 Station B, Nashville, TN 37235, USA}
\altaffiltext{2}{Fisk University, Department of Physics, 
1000 17th Ave. N, Nashville, TN 37208, USA}
\altaffiltext{3}{Department of Physics, Lehigh University, 16 Memorial Drive East, Bethlehem, PA, 18015}

\begin{abstract}
Most extrasolar planets have been detected by their influence on their parent star, 
typically either gravitationally (the Doppler method) or by the small dip in 
brightness as the planet blocks a portion of the star (the transit method).  
Therefore, the accuracy with which we know the masses and radii of extrasolar 
planets depends directly on how well we know those of the stars, the latter usually 
determined from the measured stellar surface gravity, \logg.  Recent work has 
demonstrated that the short-timescale brightness variations 
(``flicker") of stars can be used 
to measure \logg\ to a high accuracy of 
$\sim$0.1--0.2 dex \citep{bastien13}. Here, we use flicker
measurements of 289
bright (Kepmag$<$13) candidate 
planet-hosting stars with \teff=4500--6650~K
to re-assess the stellar parameters and 
determine the resulting impact on derived planet properties.  This re-assessment 
reveals that for the brightest planet-host stars, an astrophysical bias exists 
that contaminates the stellar sample with evolved stars: nearly 50\% of the bright 
planet-host stars are subgiants. As a result, the stellar radii, and hence the 
radii of the planets orbiting these stars, are 
on average 20--30\% larger than previous 
measurements had suggested.
\end{abstract}


\section{INTRODUCTION\label{intro}}
NASA's {\it Kepler} Mission \citep{borucki10}, which monitored the brightnesses 
of $>$150\,000 stars, has uncovered $>$3\,000 transiting planetary candidates 
\citep{batalha13,burke14}.  In order to measure the effective temperatures 
(\teff) and surface gravities (\logg) of 
this large number of stars---with the core purpose 
of quickly identifying as many 
likely dwarf stars as possible and to screen out as many evolved stars as 
possible to achieve the primary mission goals---the 
mission has of necessity relied on broad-band photometry. 
This is the most efficient method
for estimating stellar parameters, but with uncertainties 
in \logg\ of 0.35--0.6 dex \citep{brown11}.  
Through community followup observations, a number
of the {\it Kepler} targets have been observed spectroscopically, 
typically reducing the reported uncertainty in \logg\ to $\sim$0.1--0.2 dex 
\citep[though with the possibility of systematic offsets up to $\sim$0.4 dex;][]{torres12}.

Because such uncertainties in \logg\ translate 
into similarly large uncertainties in the derived planet radii,
many previous analyses have
attempted to mitigate stellar parameter uncertainties by imposing
priors based on theoretical stellar evolutionary tracks.
This results in a better match of the inferred \teff\ and \logg\ to the 
theoretical main sequence, but
also results in an underestimate of subgiant frequency in, e.g.,
the {\it Kepler} Input Catalog \citep[KIC;][]{brown11,huber14,everett13}.

Planets orbiting bright stars are 
of particular interest as these offer the most opportunities for 
follow-up investigation such as radial-velocity studies and in-depth spectroscopic 
analysis. However, magnitude-limited samples are generally 
not representative of the Galaxy's underlying stellar population, one of several 
biases \citep{gaidos13} that affect the {\it Kepler} transiting
planet candidates. In particular, 
magnitude-limited samples can be strongly biased toward stars that are intrinsically 
more luminous (i.e., physically larger) than the main-sequence dwarfs that 
comprise $\sim$85\% of stars in the Galaxy. Hence, it is imperative to ascertain 
the true \logg\ of the stellar hosts, the bright ones especially, as accurately 
as possible.

\citet{bastien13} demonstrated that the 
8-hr ``flicker'' ($F_8$) in the {\it Kepler} light curves can be used to measure
stellar \logg\ with an accuracy of $\sim$0.1--0.2 dex from its correlation with 
granulation power \citep{mathur11,kjeldsen11,cranmer14}. 
Thus, \logg\ determined from $F_8$ for planet-host stars can potentially 
significantly improve the inferred parameters of the planets.
Here, we use $F_8$ to refine the stellar \logg\ for the 
bright {\it Kepler} Objects of Interest (KOIs, which include both candidate 
and confirmed planets) with magnitudes of Kepmag$<$13, and we
re-examine the planet radii resulting from these revised stellar \logg.

\section{DATA AND ANALYSIS\label{data}}

\subsection{KOI Target Selection}
We draw our bright KOI sample from the NASA Exoplanet Archive 
\citep[NEA;][]{akeson13} accessed on 07 Jan 2014. 
We restrict the sample to stars with 6650~K$>$\teff$>$4500~K, the 
\teff\ range for which $F_8$ is calibrated.  We exclude 28 stars with 
overall range of photometric variability
$>$10~ppt (parts per thousand), as phenomena in the light curves 
of such chromospherically 
active stars can boost the measured $F_8$ and thus result in an erroneous 
$F_8$-based \logg.  These excluded stars (10\% of the sample) are cooler 
than average for the overall sample, as expected given their large variability.
Our sample after applying these cuts contains 289 stars (407 KOIs).  
We compare the $F_8$-based \logg\ with values from 
the recently published {\it Kepler} Stellar Properties catalog \citep{huber14}.  
Many of these values were obtained from the original KIC, whose core 
purpose was to ensure that as many dwarfs as possible were included among the 
{\it Kepler} targets at the risk of suffering contamination from some more 
evolved stars.

\begin{figure}[ht]
\includegraphics[width=8.4 cm]{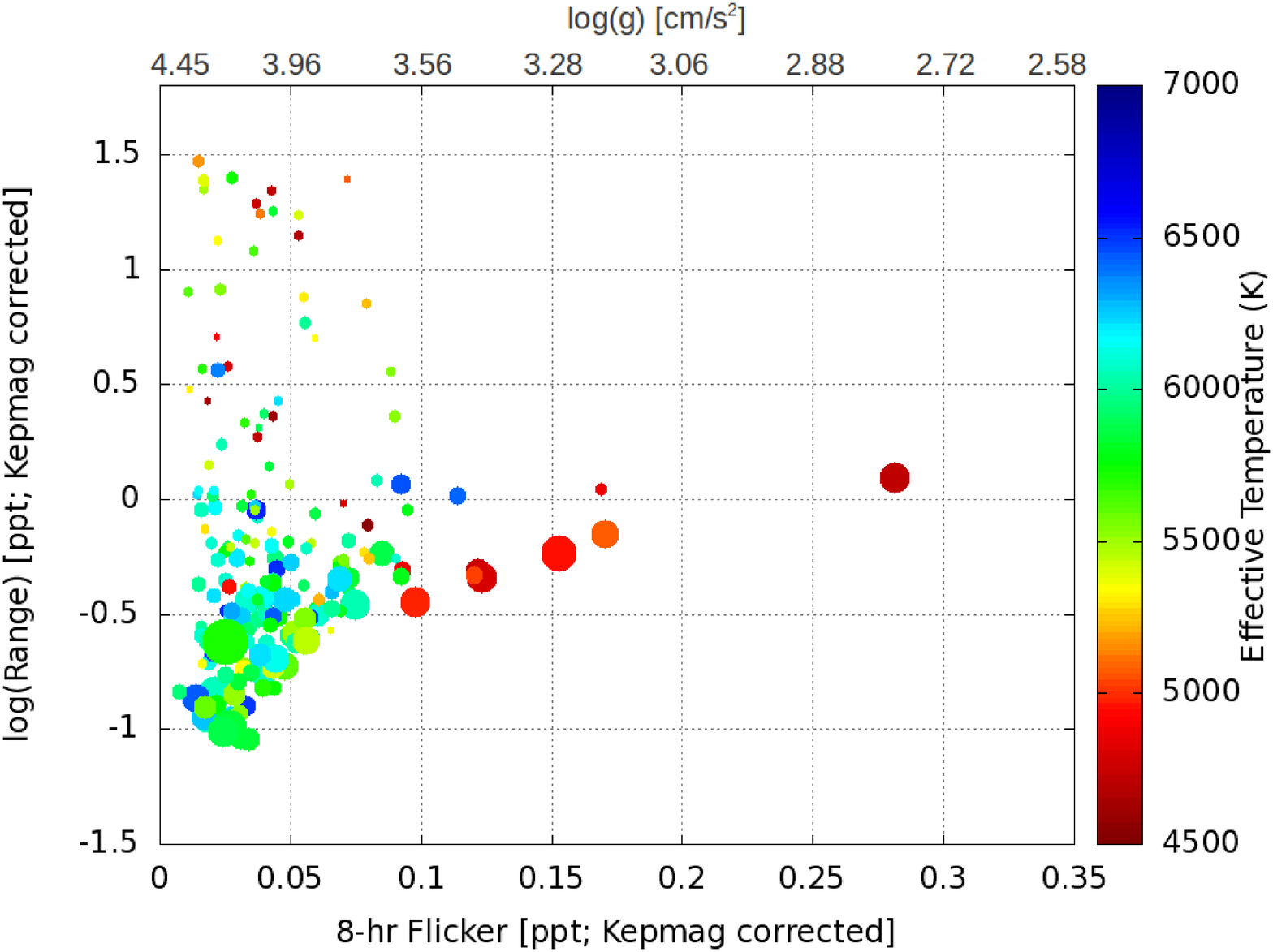}
\includegraphics[width=8.4 cm]{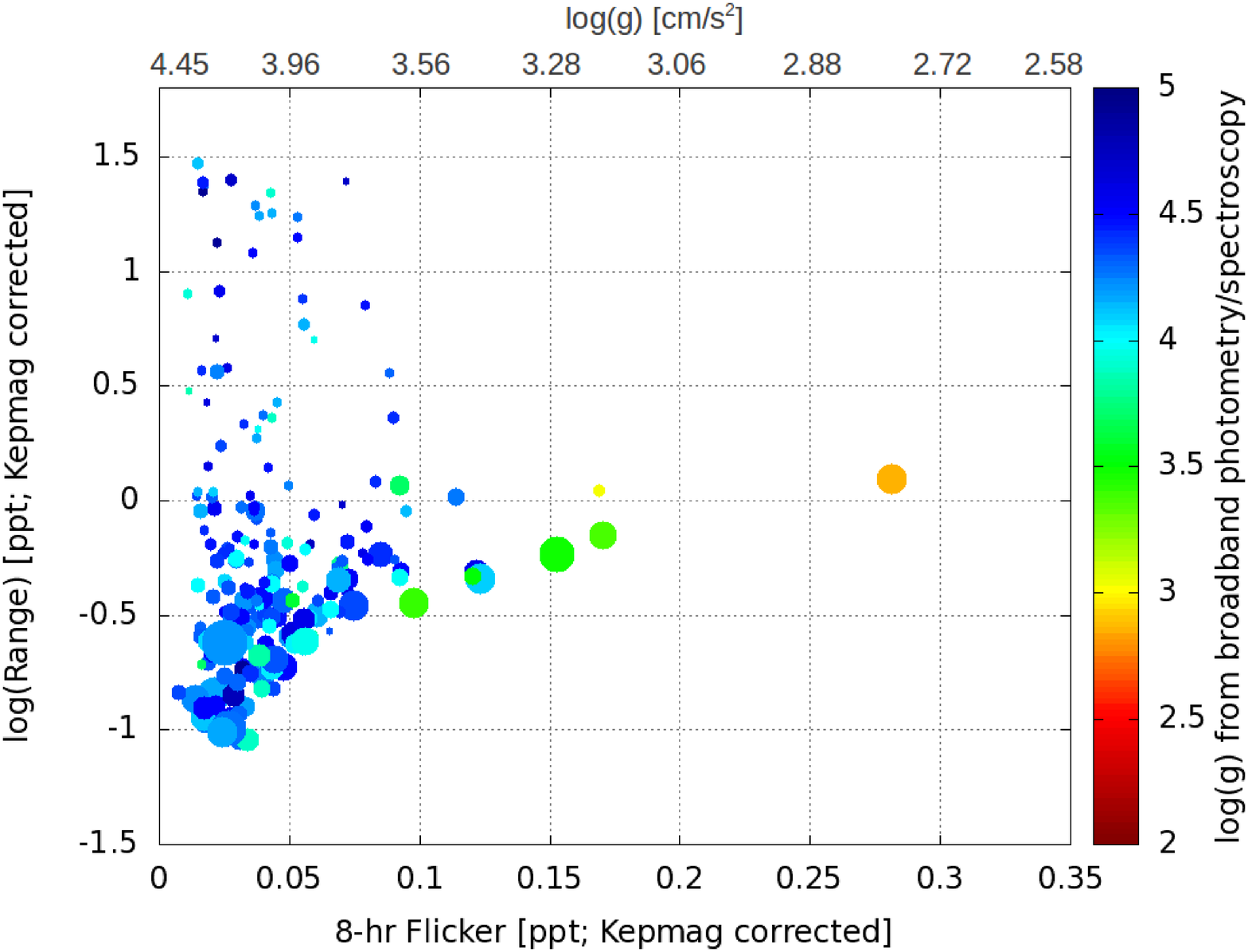}
\includegraphics[width=8.4 cm]{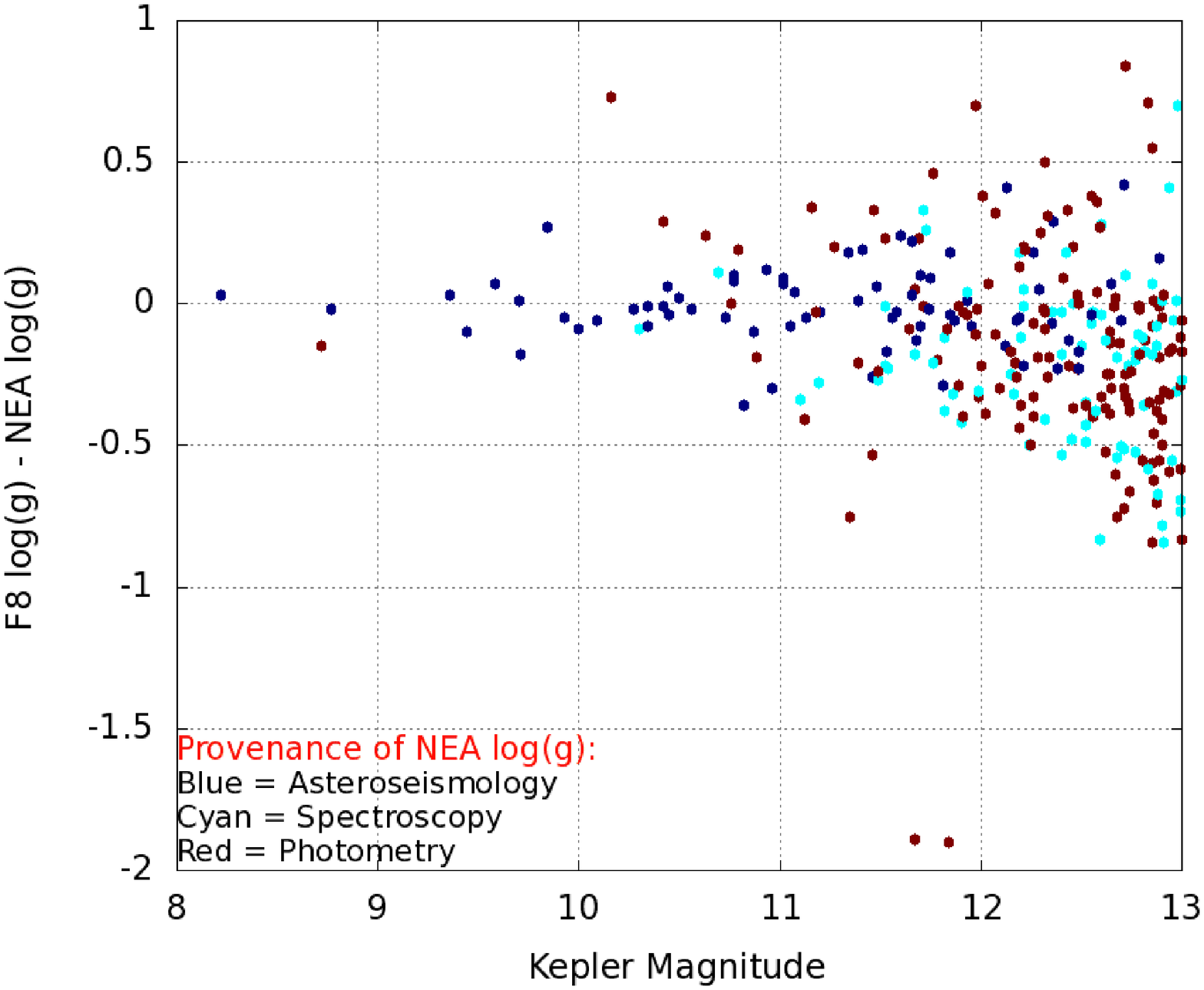}
\caption{\label{fig:evoldiag}
{\it Evolutionary states (\logg) of KOI host stars.}  
{\it Top}: On the abscissa is 8-h flicker ($F_8$) and its corresponding \logg\ scale
\citep{bastien13}.  The amplitude of the photometric variations, 
R$_{\rm var}$, is on the 
ordinate, the number of zero crossings ($X_0$) is represented as symbol size,
and color-coded by \teff\ \citep{huber14}. A 
large fraction of KOIs with $F_8$ less than 0.05~ppt 
(i.e., dwarfs) have R$_{\rm var}$$<$1~ppt, partially reflecting the 
preference for magnetically quiet stars among the KOIs. 
{\it Middle}: Color-coded by \logg\ previously determined from broadband 
photometry/spectroscopy (NEA), the NEA \logg\ indicate 
the stars are mostly dwarfs, while $F_8$-based \logg\ shows many of them are 
significantly more evolved.
{\it Bottom}: Comparison of $F_8$-based and NEA \logg\ vs.\
Kepler magnitude, color-coded by method used to derive the NEA \logg.
{\it A color version of this figure is included in the electronic journal.}}
\end{figure}

In Figure~\ref{fig:evoldiag}, we represent \citep{burger13} 
these 289 planet host stars on the photometric variability 
evolutionary diagram introduced by \citet{bastien13}.  This 
diagram traces the evolution of Sun-like stars with three simple measures of 
their brightness variations \citep{basri11}: range (R$_{\rm var}$), number of zero 
crossings ($X_0$), and root mean square (r.m.s.) on timescales shorter than 8 hours 
(8-hr ``flicker" or $F_8$; see Section~\ref{sec:f8}).  
Most of the KOIs orbit stars with R$_{\rm var}$$<$1~ppt, 
reflecting the preference for searches around magnetically quiet stars, and 
$F_8$-based \logg\ greater than 3.5 (indicative of dwarfs or subgiants).  
Some of the stars lie on the ``flicker floor'' and have \logg\ as 
low as $\sim$2.7, making them evolved giants.

\subsection{Measurement of Stellar Surface Gravity\label{sec:f8}}

To derive the $F_8$-based \logg\ of a star, we take all of the available PDC-MAP 
{\it Kepler} light curves and remove all known planet candidate transits using 
publicly available (NEA) orbital parameters.  To remove remaining 
outlying points (flares, data artifacts, etc.), we apply a 2.5~sigma 
clipping to the resultant light curves.  
We then calculate $F_8$-based \logg\ for all light curves of the star 
following the methodology of \citet{bastien13} and take the median of the 
quarter-by-quarter $F_8$ as our robust estimate of the $F_8$-based \logg.  

In general, the quarter-to-quarter 
variations in the $F_8$-based \logg\ are $<$0.1~dex.  However, 
the true accuracy of the $F_8$-based
\logg\ is a mild function of R$_{\rm var}$
\citep{bastien13}.
Stars with R$_{\rm var}$$<$1 ppt show a constant scatter 
relative to the asteroseismic calibration sample of 
$\sim$0.1 dex in \logg, whereas stars with R$_{\rm var}$$>$1 ppt show a 
slightly increased scatter of $\sim$0.15 dex. 
Therefore, we assign an 
uncertainty of either 0.1 or 0.15 dex to the $F_8$-based \logg\ depending on 
whether the total R$_{\rm var}$ is less than or greater than 1 ppt.

Additionally, the asteroseismic calibration sample used in \citet{bastien13} only included 
stars brighter than Kepmag$=$12; the asteroseismic set in \citet{huber14} 
extends to Kepmag$\sim$13 and reveals that the uncertainty in $F_8$-based \logg\ 
increases to $\sim$0.2 dex for stars fainter than Kepmag$=$12
(Fig.~\ref{fig:evoldiag}c).  We therefore
increase the assigned $F_8$-based \logg\ uncertainty to
0.2 dex for Kepmag$=$12--13.
Note that the comparison to the asteroseismic sample indicates that
the $F_8$-based \logg\ start to become unreliable for highly evolved
giants with \logg$\lesssim$2.7 \citep{bastien13}; thus
in addition to restricting our analysis to the \teff\ range of
4500--6650~K for which $F_8$ is calibrated,
we also disregard stars with \logg$<$2.7.

\subsection{Determination of Planet Properties}
To determine the planetary radius for each
KOI, we begin with the NEA-reported planet-to-star radius ratios. For the stellar 
radii, we use the empirical relationship between stellar radius, \teff, \logg, and 
metallicity \citep{torres10}, where we use 
NEA metallicities and \teff\ \citep{huber14} together with our newly determined
$F_8$-based \logg\ values. \citet{huber14} used isochrone fitting to derive 
the NEA stellar radii from \teff, metallicity, and \logg. 
\citet{torres10} showed that stellar masses and radii resulting from the empirical 
relations agree with those of model isochrones to within $\sim$5\%. 
This difference in method for determining stellar radii, therefore, 
does not change the core results of this work.

When multiple parameter estimates are available 
for a given star, we favor asteroseismic parameters, followed in 
priority order by spectroscopy, transit analyses, and lastly 
broadband photometry which often includes original KIC values (note that the resulting 
sample contains no objects with transit-derived properties).

The result is a sample of stellar parameters that is necessarily heterogeneous but 
whose average accuracy in \logg\ we expect surpasses the original KIC.  
We also note that the spectroscopic sample is itself heterogeneous in quality 
and signal-to-noise. An assessment of the individual spectra and their analysis is 
beyond the scope of this paper, and we consider the spectroscopic results together 
as an ensemble.  We provide the final stellar parameters that we use, including our 
$F_8$-based \logg\ and the NEA parameters, in Table~\ref{tab:data}.

\begin{deluxetable}{rcrrcrc}[ht]
\tablecaption{\label{tab:data}
Stellar parameters for study sample}
\tablecolumns{7}
\tablewidth{0pt}
\tablehead{
\colhead{KIC} & \colhead{Kepmag} & \colhead{\teff} & 
\colhead{\logg} & \colhead{\logg} &
\colhead{R$_{\rm var}$} & \colhead{Source\tablenotemark{a}} \\
\colhead{} & \colhead{} & \colhead{[K]} & \colhead{($F_8$)} &
\colhead{(NEA)} & \colhead{[ppt]} & \colhead{}
}
\startdata
 3109550  &  12.21  &  5449  &  4.23  &  4.03  &  4.64  &  3 \\ 
 3114811  &  12.81  &  6350  &  3.80  &  3.97  &  0.48  &  2 \\ 
 3240159  &  12.26  &  6413  &  3.95  &  4.28  &  2.40  &  3 \\ 
 3328080  &  12.99  &  5702  &  3.68  &  4.26  &  0.36  &  3 \\ 
 3425851  &  10.56  &  6343  &  4.24  &  4.26  &  0.37  &  1 \\
\enddata
\tablenotetext{a}{Flag for source of NEA \logg: 1 = asteroseismic (72 
stars), 2 = spectroscopic (78 stars), 3 = photometric (139 stars)}
\tablecomments{A portion of the table is shown for guidance regarding
content and format. The full table is available electronically.}
\end{deluxetable}

\section{RESULTS\label{results}}

Comparing the \logg\ previously estimated from broadband photometry/spectroscopy 
versus that newly measured via $F_8$
(Fig.~\ref{fig:evoldiag}c), we find the 
$F_8$-based \logg\ to be systematically 
lower (i.e., more subgiant-like), with a median difference of 
$\sim$0.2 dex (r.m.s.\ of 0.3 dex for spectroscopy, 
0.4 dex for photometry).  In contrast, the subset of 
the stars with \logg\ determined asteroseismically agrees with the $F_8$-based 
\logg\ to 0.02~dex in the median (r.m.s.\ of 0.15 dex), consistent 
with the expected accuracy of the $F_8$-based \logg\ (Sec.~\ref{sec:f8}). 
While the scatter of 0.3--0.4 dex in the spectroscopic/photometric \logg\ is 
consistent with the expected precision of photometry, 
it is large compared to that expected of spectroscopy.  The 
asteroseismic and $F_8$-based \logg\ together appear to indicate a significant 
overestimate of the NEA \logg\ for the bright KOI stars
(the overestimate increasing to fainter magnitudes; Fig.~\ref{fig:evoldiag}c). 
We stress that we also find this overestimate for a number of stars whose
NEA \logg\ is spectroscopically derived. 
This result has also been reported in asteroseismic studies, and
may be due to biases in spectroscopic analyses
that impact the \logg\ determined for giants and subgiants 
\citep[see, e.g.,][for a discussion]{huber13}.

\begin{figure}[ht]
\includegraphics[width=8.4 cm]{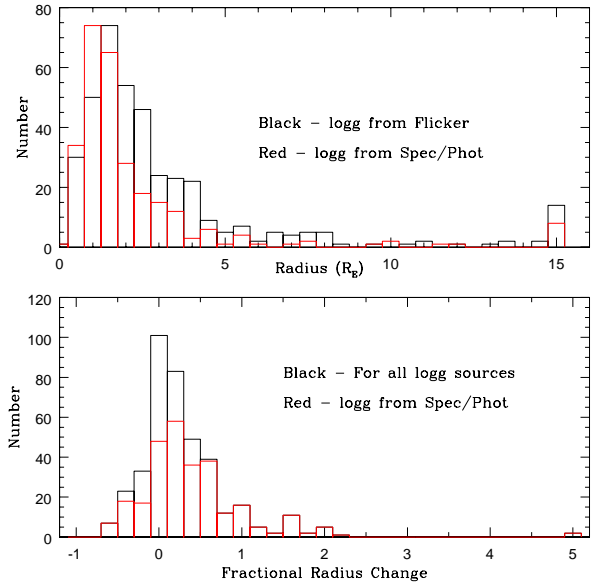}
\caption{\label{fig:batalha}
{\it Distribution of planet candidate radii, according to \logg\ from 
$F_8$.}  
{\it Top}: 
Distribution of planet orbital period against exoplanet radius.  The arrows 
depict how planet radii shift when they have \logg\ 
derived from $F_8$ (arrow head) vs.\ \logg\ estimated from broad-band 
photometry/spectroscopy (arrow tail).  
{\it Middle}:  The distribution of exoplanetary radii with stellar \logg\ obtained 
from $F_8$ (black curve) and broad-band photometry/spectroscopy (red curve).  
{\it Bottom}:  Fractional change in planet radius between $F_8$-based \logg\ 
and photometry/spectroscopy-based \logg.  Using $F_8$-based stellar \logg\ 
results in exoplanet radii that are $\sim$20--30\% larger than suggested by 
broad-band photometric/spectroscopic \logg.
{\it A color version of this figure is included in the electronic journal.}}
\end{figure}

The key effect of including $F_8$-derived \logg\ is to significantly 
increase the median radius of the bright KOIs 
(Figure~\ref{fig:batalha}).  We find that the 
median KOI radius is larger by 20--30\% compared to that inferred using the 
\logg\ previously estimated from broadband photometry or spectroscopy, 
though a number of objects show a more modest or even negative change in radius 
(Figure~\ref{fig:batalha}b). 

To compare our results with those expected based on the underlying stellar 
population, given the magnitude-limited nature of the sample, we 
simulated the {\it Kepler} field using the TRILEGAL Galactic population synthesis 
model \citep{girardi05}. We used the default TRILEGAL model parameters, for 
a 1~deg$^2$ line-of-sight toward the center of the {\it Kepler} field, and we include only
simulated stars down to Kepmag$<$13 and with 
6650~K$>$\teff$>$4500~K, as for the KOI sample.

\begin{figure}[ht]
\includegraphics[width=8.4 cm]{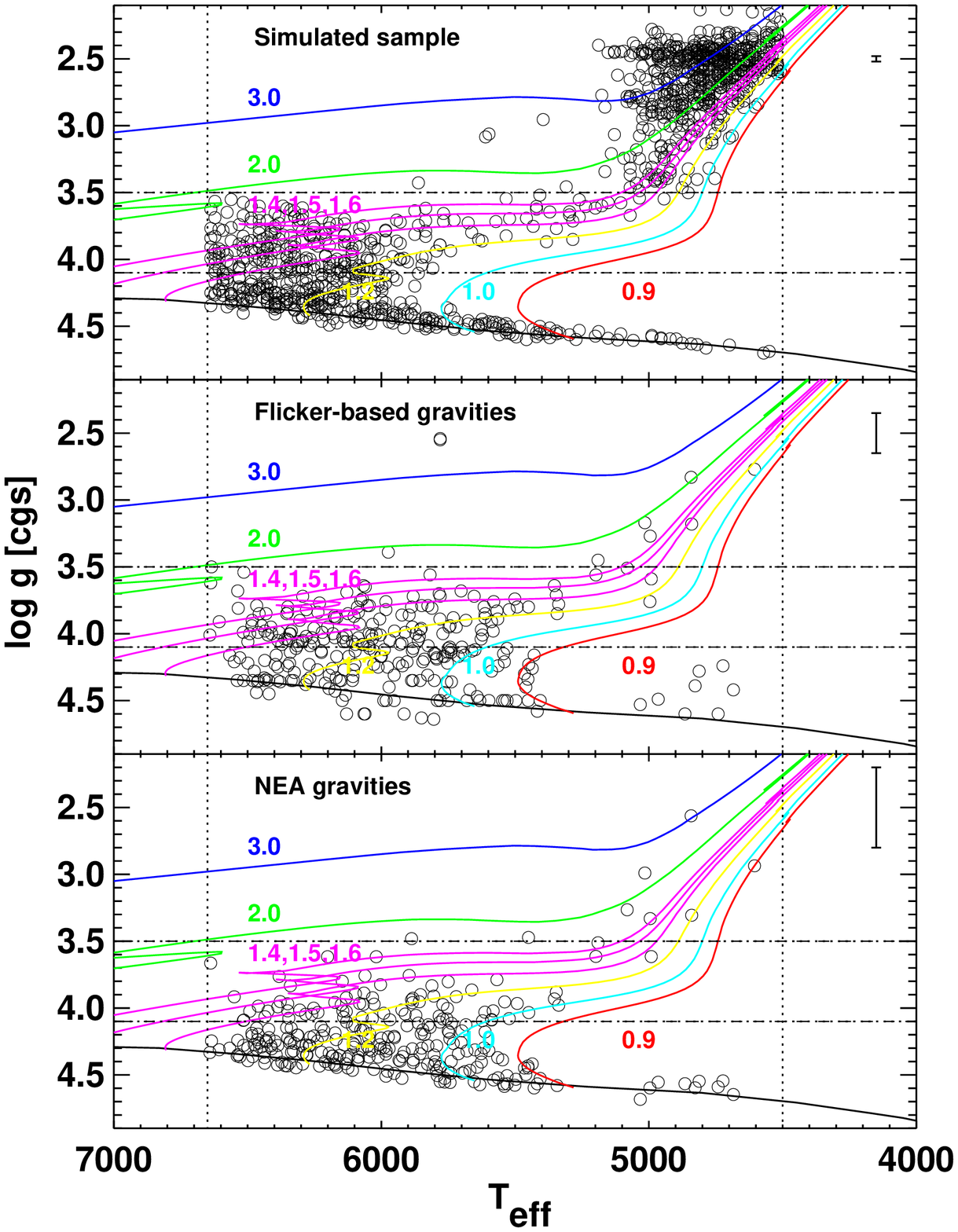}
\caption{\label{fig:trilegal_hr}
{\it H-R diagram of KOI host stars} 
with \logg\ derived from $F_8$ ({\it middle}) and broadband photometry/spectroscopy 
({\it bottom}), and as predicted by a TRILEGAL \citep{girardi05} simulation
({\it top}).  Colored curves represent the theoretical evolutionary tracks 
(masses labeled in \msun). Vertical lines demarcate the 
range of stellar \teff\ considered in this study. The horizontal 
lines demarcate the range of \logg\ for subgiants 
(3.5$<$\logg$<$4.1). A representative error bar on \logg\ for each stellar 
sample is in the upper right of each panel.  We find that the $F_8$-based 
\logg\ distribution more closely matches expectation than 
previous \logg\ measurements, particularly in the subgiant domain, 
perhaps because $F_8$ involves no main-sequence prior on
the $F_8$-based \logg\ values (see text).
{\it A color version of this figure appears in the electronic journal.}}
\end{figure}

Fig.~\ref{fig:trilegal_hr}a shows the H-R diagram 
of the simulated population compared with 
the actual KOI sample using $F_8$-based (Fig.~\ref{fig:trilegal_hr}b) and NEA 
(Fig.~\ref{fig:trilegal_hr}c) \logg\ values.
We retain the full set of $\sim$1200 stars 
produced by the TRILEGAL simulation to visually preserve the 
detail of the parameters; the actual sample with $\sim$300
stars necessarily appears sparser. 
By construction, the simulated sample closely traces the theoretical 
evolutionary tracks, with both a tight main-sequence population along the bottom
and a large red giant population at upper right being most prominent, as
expected for a magnitude-limited population 
including a mix of stellar masses 
and ages. For stellar masses $\gtrsim$1~\msun, the simulated sample also
includes a large population of modestly evolved subgiants
with masses $\sim$1--2~\msun, forming a thick but well defined horizontal band 
with 3.5$<$\logg$<$4.1.

By comparison, the observed H-R diagram 
({\it middle} or {\it bottom} panel) 
lacks the highly evolved red giants (\logg$<$3)
present in the simulated sample, the result of their 
systematic removal 
by the {\it Kepler} mission \citep{batalha10}.
Additionally, there is a noticeable dearth of mid-F type stars with
\teff$\gtrsim$6500~K, as well as late-K type stars with 
\teff$\lesssim$4700~K, among the observed KOI sample, likely
the result of the {\it Kepler} target selection process that
strongly favored late-F to early-K type dwarfs for Kepmag$<$13
\citep{batalha10}. 
For stars with 6500~K$>$\teff$>$4700~K and Kepmag$<$13, 
the {\it Kepler} target sample should be representative of the 
field for all but evolved giants with \logg$<$3.5 \citep{batalha10}.

More importantly, 
in the H-R diagram using \logg\ previously estimated 
from broadband photometry/spectroscopy ({\it bottom} panel)
the KOI sample overall follows the
main sequence very closely, with few apparent subgiants with \logg$<$4.1.
With the exception of a few cool stars beginning their ascent up the red giant
branch (at \teff$\sim$5000~K and \logg$\sim$3.5), there are apparently 
very few of the warmer, $\sim$1--2~\msun\ subgiants with \logg\ as low as 3.5 
that are expected from the simulated sample.

In contrast, the H-R diagram using $F_8$-based \logg\
matches the simulated stellar population more closely. 
In particular, the subgiant population predicted by the simulated
sample is more clearly present. Indeed, the $F_8$-based
\logg\ values extend down to, but cleanly truncate at, 
\logg$\approx$3.5 for \teff$\gtrsim$5200 K, just as in the 
simulated population. At higher \logg, the $F_8$-based \logg\
also trace the slope of the main sequence, with a scatter
generally within $\sim$1$\sigma$ of that expected for the 
$F_8$-based \logg\ (0.1--0.2 dex). Note that the $F_8$-based
\logg\ are not forced to match isochrones and more generally 
have no priors applied to the resulting \logg's.

At the same time, the $F_8$-based H-R 
diagram does not perfectly match the simulated sample. For example, for stellar 
masses $\lesssim$1 \msun, the $F_8$-based H-R diagram includes a few stars that
appear elevated by 1--2$\sigma$ relative to the main sequence (e.g.,
at \teff$\sim$4800~K and \logg$\sim$4.3). Since stars less massive than 
$\sim$0.9 \msun\ cannot be evolved, these stars should be firmly on
the main sequence. The NEA \logg\ values in this region of the H-R diagram
appear better behaved, a consequence of the prior that is imposed in most
photometric/spectroscopic \logg\ analyses to force the stellar parameters
to match theoretical isochrones \citep[e.g.,][]{huber14}.
The $F_8$ method imposes no prior on the \logg\ values, and so it is not
surprising that for some stars the inferred \logg\ may scatter by 1--2$\sigma$
into ``forbidden" regions of the H-R diagram. 
This may also be partially a result of the fact that $F_8$ is
fundamentally calibrated to the
{\it Kepler} asteroseismic sample and to the Sun \citep{bastien13},
such that $F_8$-based \logg$\gtrsim$4.5 constitute an 
extrapolation from that calibration.
However, we also cannot rule out the possibility that the true radii of the
low-mass stars are in fact larger than predicted by theoretical main-sequence
models, as recent interferometric observations of low-mass planet-host stars have 
found the stellar radii to frequently be larger than previously thought
\citep[see][]{vonbraun11,vonbraun12,boyajian12,vonbraun14}.
For example, of the six stars with masses $\sim$0.8--0.9~\msun\ included
in these studies, four of them (55\,Cnc, 61\,Vir, rho\,CrB, HD\,1461) are 
similarly elevated above the theoretically expected main sequence 
\citep{vonbraun14} for reasons that are not yet clear but which may include
the effects of magnetic activity \citep[e.g.,][]{stassun12}. 
In any event, for the bright KOIs considered here with very few such 
low-mass stars, this issue affects 1--2\% of the sample.

\begin{figure}[ht]
\includegraphics[width=8.4 cm]{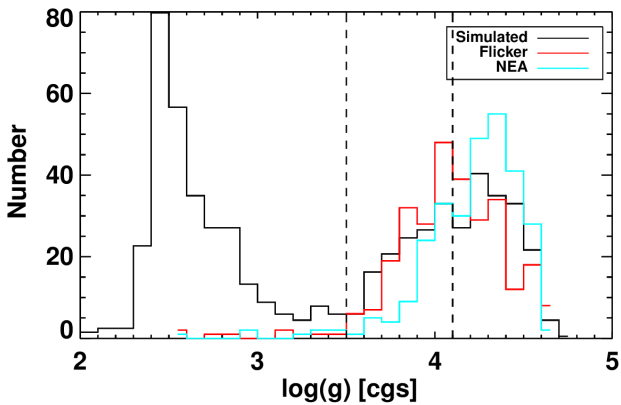}
\caption{\label{fig:trilegal}
{\it Distributions of \logg} for
simulated (black) and KOI host stars with 
$F_8$-based \logg\ (red) and broadband photometry/spectroscopy-based \logg\
(cyan). We limit the \teff\ range here to 4700--6500~K, for which the
{\it Kepler} targets should be representative of the field
\citep{batalha10}. Vertical lines indicate
the range of \logg\ corresponding to subgiants.
{\it A color version of this figure appears in the electronic journal.}}
\end{figure}

Figure~\ref{fig:trilegal} shows these \logg\ comparisons directly.
Here we limit the \teff\ range to 4700--6500~K for which the observed
{\it Kepler} targets should be representative of the field and
therefore most directly comparable to the simulated population (see above).
Again, the very large population of red giants with \logg$<$3.5 seen in the 
simulated sample is conspicuously missing in the actual planet 
host-star sample.
Thus we compare the distributions only for \logg$>$3.5, and we
normalize the histogram of the simulated sample by the number
of observed stars with \logg$>$3.5. A two-sided 
K-S test gives a probability of 0.01\% that the NEA \logg\ and the 
\logg\ from the simulated stellar population are drawn from the same parent sample, 
whereas a K-S test gives a probability of 16\% that the simulated and $F_8$-based 
\logg\ samples are drawn from the same parent sample. 
The $F_8$-based \logg\ show the poorest match to the simulated distribution
at the highest \logg, corresponding to the low-mass main-sequence dwarfs that the 
simulation assumes to be unevolved but for which the $F_8$-based \logg\ indicate
larger radii in some cases
(see above). However, the $F_8$-based \logg\ are a better match overall
to the simulated sample, in particular in reproducing the expected population
of subgiants.
Specifically, 48\% of the stars have $F_8$-based 
\logg\ values indicative of modestly evolved subgiants (3.5$<$\logg$<$4.1), 
whereas previously estimated \logg\ values had indicated that only 27\% are 
subgiants. In comparison, 
44\% of the simulated sample are subgiants, in good agreement with the 48\% 
inferred from the $F_8$ analysis.

\section{CONCLUSIONS}

In this work, we used the granulation ``flicker" ($F_8$) of bright 
{\it Kepler} planet candidate host stars to measure improved stellar \logg\ and 
to thereby redetermine the planet radii.  
Comparing the $F_8$-based \logg\ values with those previously published
in the NEA, the latter representing a
heterogeneous mix of broad-band photometric, spectroscopic, and 
asteroseismic methods, indicates that the stellar, and hence planetary, radii 
are on average 20--30\% larger than suggested by the previous estimates.  

The H-R diagram positions of the stars according to the $F_8$-based \logg\
in general appear better matched to the distribution from the simulated Galactic
population of bright stars (Kepmag$<$13) in the {\it Kepler} field, 
especially the presence of a significant population of subgiants. 
However, for the few very low-mass stars ($\lesssim$0.9 \msun) in the
sample, the \logg\ values from
photometry/spectroscopy appear to better match expectations as these are 
generally forced to match the main sequence.
Whether this is a failing of the extrapolation of the $F_8$-\logg\
relation to \logg$\gtrsim$4.5, or a manifestation of larger-than-predicted
stellar radii for low-mass K and M stars as observed interferometrically
\citep[e.g.,][]{vonbraun14,boyajian12}, remains to be determined.
The performance of $F_8$-based \logg\ for very low-mass stars
will be an important area for continued refinement of the $F_8$
technique, including its application in contexts such as 
asterodensity profiling \citep{kipping14}.

Most importantly, for a magnitude-limited sample such as that considered here, 
modestly evolved subgiants represent a large fraction of the population. 
Methods that apply a strong prior favoring main-sequence dwarf \logg\ will 
systematically overestimate \logg\ for such a sample, and in turn systematically 
underestimate the planet radii, particularly among the brightest stellar hosts. 
Our finding that broadband photometric and spectroscopic methods yield 
systematically larger stellar \logg\ than asteroseismic or $F_8$-based 
methods---especially among subgiants---is consistent with previous reports 
\citep{huber13}
but now demonstrated for a much larger sample. This bias directly impacts our 
understanding of the true distribution of exoplanetary radii, especially for the 
scientifically valuable bright systems.  The results reported herein also demonstrate 
that one cannot ignore the magnitude-limited nature of the stellar samples when 
inferring their ensemble properties.

\acknowledgments
This research has made use of the NASA Exoplanet Archive, which is operated by the 
California Institute of Technology, under contract with the National Aeronautics and 
Space Administration under the Exoplanet Exploration Program.  We acknowledge 
helpful discussions with J.\ Eastman.  We acknowledge NSF PAARE grant AST-0849736,
and NASA Harriet Jenkins and Vanderbilt Provost Graduate Fellowships.

\end{document}